\documentclass[preprintnumbers,floatfix,twocolumn,superscriptaddress,a4paper,nofootinbib]{revtex4}

\usepackage{epsfig,graphicx}
\usepackage{amsmath,amsfonts,amssymb} 
\usepackage{color}
\usepackage{slashed}
\usepackage{hyperref}

\usepackage{soul}

\usepackage[capitalise, english]{cleveref}

\usepackage{graphicx}% Include figure files
\usepackage{dcolumn}% Align table columns on decimal point
\usepackage{bm}% bold math
\def\met{\slash\hspace*{-1.5ex}E_{T}}

\def\lsim{\raise0.3ex\hbox{$\;<$\kern-0.75em\raise-1.1ex\hbox{$\sim\;$}}}
\def\gsim{\raise0.3ex\hbox{$\;>$\kern-0.75em\raise-1.1ex\hbox{$\sim\;$}}}

\allowdisplaybreaks

\definecolor{vmlorange}{rgb}{1.0, 0.49, 0.0}
\definecolor{mkgreen}{rgb}{0.2,.70,.3}

\begin{document}

\title{Probing the Electroweakino Sector of General Gauge Mediation at the LHC}% Force line breaks with \\
\preprint{BONN-TH-2017-04}

\newcommand{\AddrBonn}{%
Bethe Center for Theoretical Physics \& Physikalisches Institut der 
Universit\"at Bonn,\\ Nu{\ss}allee 12, 
 53115 Bonn, Germany
}

\newcommand{\AddrSA}{
National Institute for Theoretical Physics,\\
School of Physics and Mandelstam Institute for Theoretical Physics,\\
University of the Witwatersrand, Johannesburg,
Wits 2050, South Africa  
}

\author{Jong Soo Kim} \email{jongsoo.kim@tu-dortmund.de}
\affiliation{\AddrBonn}\affiliation{\AddrSA}%

\author{Manuel E. Krauss} \email{mkrauss@th.physik.uni-bonn.de}\affiliation{\AddrBonn}
\author{V\'ictor Mart\'in Lozano} \email{lozano@physik.uni-bonn.de}
\affiliation{\AddrBonn}


\begin{abstract}
We consider pair-production of electroweakinos promptly decaying to light gravitinos in 
general gauge mediation scenarios within the minimal supersymmetric standard model. Typically the presence of photons and missing transverse momentum is the key signature for this kind of scenarios. We highlight where LHC analyses which have originally been designed to probe different scenarios 
provide complementary constraints with respect to the dedicated searches and we present the constraints on the parameter space.
\end{abstract}

\maketitle

%\tableofcontents
% -----------------------------------------------------------
% ARTICLE
% -----------------------------------------------------------

\section{Introduction}
\label{sec:introduction}

After the discovery of a Higgs boson at the LHC in 2012 \cite{ATLAS:2012ae,Chatrchyan:2012tx}, theorist's expectations for signs of new physics at the TeV scale have so far not been met. Instead, many models beyond the Standard Model (BSM) such as models of supersymmetry (SUSY) where the abundant production of squarks and gluinos has been long anticipated,  suffer from severe constraints because of the perpetual null results. In a combined global fit \cite{Bechtle:2015nua}, those already led to the first 90\% confidence level (C.L.) exclusion of the constrained minimal supersymmetric standard model, which assumes Planck-scale mediated SUSY breaking \cite{Nilles:1983ge,Drees:2004jm}. An attractive alternative way of breaking SUSY is gauge-mediated SUSY breaking (GMSB) \cite{Dine:1981gu,Dine:1981za,Dimopoulos:1981au,Nappi:1982hm,AlvarezGaume:1981wy,Dine:1993yw,Dine:1993qm,Dine:1994vc,Dine:1995ag} where SUSY is broken in a hidden sector and the breaking is transmitted to the visible world by so-called messenger-fields. 
Being charged under the gauge group of the model, those couple to the matter fields through gauge interactions. Because of the flavour-blindness of gauge mediation, there is no SUSY flavour problem.
The mass scale of SUSY breaking as well as of the messenger fields is typically considerably smaller than the Planck scale. Consequently, the SUSY-breaking mass of the longitudinal mode of the graviton partner, the gravitino, is so small that it is the lightest supersymmetric particle (LSP).

If $R$-parity is conserved, the gravitino is a viable candidate for dark matter (DM)~\cite{Pagels:1981ke,Khlopov:1984pf}. In general thermal scenarios, the gravitino could thermalise giving the correct relic abundance for masses below a keV, corresponding to warm DM. However, in this scenario one should be aware of the impact of such a light candidate in astrophysical and cosmological observations~\cite{Viel:2005qj}. The lower mass bound is avoided if the gravitino has never been in thermal equilibrium
which is possible for low reheating temperatures \cite{Choi:1999xm}, e.g. obtained by gaugino scattering~\cite{Moroi:1993mb}. Furthermore, in Ref.~\cite{Cheung:2011nn} it was pointed out that the gravitino can be produced through the \textit{freeze-in} mechanism, resulting in the correct relic abundance for masses below 10\,GeV. Recently, it has been shown that this scenario can also be fulfilled up to large values of the SUSY breaking scale~\cite{Benakli:2017whb}.

In the minimal gauge mediation models~\footnote{For an overview, see Refs.~\cite{Giudice:1998bp,Martin:1996zb} and references therein.}, the masses of the SUSY scalars and of the fermionic partners of the gauge bosons (gauginos) are tightly connected. 
As the Higgs mass constraint requires masses of the stop squarks in the multi-TeV range \cite{Draper:2011aa,Ajaib:2012vc}, this means that the hope for discovering this kind of models at the LHC is small.
However, in more general scenarios of gauge mediation which also encompass models with direct mediation, this statement is not true any more. These models are referred to  as general gauge mediation (GGM) \cite{Meade:2008wd,Cheung:2007es}.
A comprehensive analysis of GGM scenarios after imposing the Higgs mass constraint which also takes into account different scenarios for the next-to-lightest supersymmetric particle (NLSP) is for instance provided in Refs.~\cite{Knapen:2015qba,Knapen:2016exe}.

Refs.~\cite{Ambrosanio:1996jn,Feng:1997zr,Stump:1996wd,Dimopoulos:1996va,Baer:2000pe,Kim:2011sv,Hiller:2009ii} investigated the discovery potential of models based on gauge mediation with light gravitinos at LEP, Tevatron and the LHC. In this letter, we shall focus on the case where the masses of the SUSY partners of the electroweak gauge and Higgs bosons reside below a TeV whereas the rest of the spectrum is considerably heavier. In that case, coloured production does not lead to any discovery prospects but one has to consider electroweak production. Analyses and interpretations in that direction have for instance been done in Refs.~\cite{Aad:2015hea,ATLAS:2016hks}.
In that respect, Ref.~\cite{Aad:2015hea} sets bounds on the electroweak production of wino pairs for both wino and bino NLSPs. Ref.~\cite{ATLAS:2016hks} is a recast of experimental analyses which look for either two photons or multi-lepton signatures, considering  weak-scale winos and higgsinos for fixed NLSP mass of 150\,GeV and decoupled binos in a GGM setting.

Lately, there has been an increased effort in the phenomenological community to provide tools which make use of actual experimental results and provide the possibility of recasting them in custom models, see e.g. Refs.~\cite{Caron:2016hib,Drees:2013wra,Conte:2012fm,Papucci:2014rja,Kraml:2013mwa,Bechtle:2017vyu}, and for recent examples of recasting Refs.~\cite{Arina:2016rbb,Deutschmann:2017bth,Beuria:2017jez,Cerna-Velazco:2017cmn}. For this project we make use of the public tool {\tt CheckMATE} \cite{Drees:2013wra,Dercks:2016npn} for which we have implemented the 
%analysis of Ref.~\cite{Aad:2015hea} and made it 
analyses of Refs.~\cite{Aad:2015hea,CMS:2017vdc} and made them  % CMS:2017mwo,
available for public usage. Using this implementation as well as all other analyses included in {\tt CheckMATE},  we expand upon the GGM analyses of Refs.~\cite{Aad:2015hea,ATLAS:2016hks} to provide a full coverage of the electroweakino sector, for instance also considering bino-higgsino NLSPs. We further show where other relevant signatures from analyses which are not specifically designed for GGM models like the mono-photon searches of Refs.~\cite{Aad:2014tda,Aaboud:2017dor} or the multilepton analysis of Ref.~\cite{ATLAS:2017uun} provide complementary bounds.

\section{General Gauge Mediation Models}
\label{sec:model}

In the minimal GMSB scenario, %\footnote{For an overview, see Refs.~\cite{Giudice:1998bp,Martin:1996zb} and references therein.} 
the spectrum of supersymmetric particles is determined by their respective quantum numbers along with the SUSY breaking scale and the
properties of the messenger sector. Typically one assumes the messenger fields to form complete multiplets under SU(5) so that gauge coupling unification is trivially preserved.
% multiplicity, representation and mass scale $M$ of the messenger fields as well as the scale of SUSY breaking.
%  Once these parameters together with the 
Once the messenger sector and the  
  ratio of the up- and down-type Higgs vevs $\tan\beta=v_u/v_d$ are fixed, the hierarchy of the spectrum is determined. In particular, the same messenger fields which fix the masses of the scalar particles also determine the gaugino spectrum, with the fixed ratio of bino, wino and gluino masses of $M_1:M_2:M_3 = g_1^2:g_2^2:g_3^2$ at the messenger scale $M$.\footnote{Note that this relation uses the GUT-normalized U(1) gauge coupling $g_1$.}

In general gauge mediation models, in turn, a more general messenger sector is considered and
interactions between matter and messenger fields are included. 
Consequently, the spectra of GGM models can differ quite significantly from minimal GMSB models, although
some features such as small softbreaking trilinear couplings ($A$-terms) and sum rules for soft SUSY-breaking masses remain \cite{Meade:2008wd}. In addition, the gravitino typically remains the LSP.

The MSSM gaugino and sfermion soft masses can be written, to leading order, as \cite{Buican:2008ws}
\begin{equation}
M_k = g_k^2 M B_k\quad {\rm and} \quad m_f^2 = \sum_{k=1}^3 g_k^4 C_k(f) A_k\,,
\end{equation} 
where $k$ labels the gauge groups, $f$ the MSSM matter representations and  $C_k (f)$ is the quadratic Casimir invariant of the scalar field $f$ with respect to the gauge group $k$. In Ref.~\cite{Buican:2008ws}, it was shown that 
one can construct weakly-coupled messenger theories in such a way that the complete parameter space of the parameters $B_k$ and $A_k$ is covered.\footnote{See also Refs.~\cite{Rajaraman:2009ga,Komargodski:2008ax} for an exploration of the GGM parameter space.} 
In addition to $A_k$ and $B_k$, the remaining free parameters at the weak scale are the higgsino mass term $\mu$ as well as $\tan \beta$. 
Consequently, the hierarchy among the electroweakinos is undetermined. Moreover, any one of them could be the NLSP with masses well below a TeV while having at the same time large enough $A_k$ such that the stops are in the multi-TeV range as required from the Higgs mass constraint in conjunction with small $A$-terms.\footnote{This problem can, however, be alleviated by certain types of messenger-matter interactions \cite{Grajek:2013ola,Byakti:2013ti,Evans:2013kxa,Craig:2012xp,Abdullah:2012tq,Albaid:2012qk,Evans:2012hg,Evans:2011bea,
Shadmi:2011hs,Kang:2012ra} or extra tree-level contributions to the Higgs mass \cite{Krauss:2013jva}.}
 If also the rest of the scalar sector and the gluinos are heavy, the first (and possibly the only) SUSY particles accessible at the LHC would be the electroweakinos.

In the following, we will investigate this scenario, treating $M_1$, $M_2$, $\mu$ and $\tan \beta$ as free parameters.
We further assume that the sfermions as well as the gluino are heavy enough to not (yet) be accessible at the LHC in the near future. 

Once produced, the NLSP will undergo the decay $\tilde \chi^0_1 \to \tilde G \gamma/Z/h$, depending on the NLSP nature.
The partial decay width into $\tilde G \gamma$ 
is for instance given by \cite{Ambrosanio:1996jn}
\begin{equation}
\Gamma_{\tilde G \gamma} = 1.12 \cdot 10^{-11}~{\rm GeV} \, \kappa_\gamma \left( \frac{m_{\tilde \chi^0_1}}{100~{\rm GeV}} \right)^5 \left( \frac{m_{\tilde G}}{1~{\rm eV}} \right)^{-2}\,,
\end{equation}
and the corresponding ratios of partial widths can be approximated as~\footnote{These relations do not hold near threshold, i.e. when for instance $m_{\tilde \chi^0_1}-(M_Z+m_{\tilde G})$ is small.} 
\begin{align}
\label{eq:partialwidths}
\frac{\Gamma_{ h \tilde G}}{\Gamma_{\gamma \tilde G}} = \frac{\kappa_h}{\kappa_\gamma} \left(1-\frac{m_h^2}{m^2_{\tilde \chi^0_1}} \right)^4\,,\quad
\frac{\Gamma_{ Z \tilde G}}{\Gamma_{\gamma \tilde G}} &= \frac{\kappa_Z}{\kappa_\gamma} \left(1-\frac{M_Z^2}{m^2_{\tilde \chi^0_1}} \right)^4 \,,
\end{align}
where
\begin{align}
\label{eq:kappas}
\nonumber \kappa_\gamma &= |N_{11} c_{W} + N_{12}s_W|^2,\,\quad\kappa_h = |N_{13}s_\alpha - N_{14} c_\alpha|^2,\,\\
\kappa_Z &= |N_{11} s_W-N_{12}c_W|^2 + \frac{1}{2} |N_{13} c_\beta-N_{14}s_\beta|^2\,.
\end{align}
Here we have used $c_\phi = \cos \phi\,,s_\phi=\sin \phi$, subscript $W$ refers to the weak mixing angle and $\alpha$ is the  mixing angle in the CP-even Higgs sector. $N_{ij}$ is the neutralino mixing matrix in the basis $(\tilde B,\tilde W^0,\tilde H^0_d,\tilde H^0_u)$.
% Complete expressions for the neutralino two-body decay into $\tilde G \gamma/Z/h$ can be found in Ref.~\cite{Giudice:1998bp}. If the NLSP is a bino, the decay $\tilde \chi^0_1 \to \gamma \tilde G$ always dominates.  
Since $\kappa_Z/\kappa_\gamma \simeq 0.3$, a bino NLSP will predominantly decay into $\tilde G \gamma$ while a pure wino NLSP mostly decays into $\tilde G Z$ as long as $(M_Z/m_{\tilde \chi^0_1})^2\ll 1$. For a pure higgsino state, the photonic mode is absent and so it decays into both $\tilde G Z$ or $\tilde G h$, with the ratio depending on the model parameters.

We will in the following restrict ourselves to prompt neutralino decays~\footnote{For discussions on long-lived neutralinos, see e.g. Refs.~\cite{Feng:2010ij,Dreiner:2011fp} and Ref.~\cite{Aad:2014gfa} for an experimental search for displaced vertices motivated by long-lived neutralino scenarios within GMSB.}
 which we assure by fixing $m_{\tilde G}=1$~eV in the remainder of this letter. In this case the gravitino is still a good candidate for DM, however it should coexist with other different candidates in order to account for the total relic density~\cite{Viel:2005qj}.
Note however that the exact gravitino mass value is irrelevant for the subsequent study due to $m_{\tilde \chi^0_1}\gg m_{\tilde G}$.

\section{Recasting the experimental searches}
\label{sec:recasting}

We generated benchmark points with the spectrum generator {\tt SPheno\,4.0.0} \cite{Porod:2003um,Porod:2011nf,Staub:2017jnp} using the scanning tool {\tt SSP1.2.3} \cite{Staub:2011dp}. 
The truth level MC events are generated with the help of {\tt Pythia\,8.219} \cite{Ball:2012cx} employing the default parton distribution function (PDF) {\tt NNPDF\,2.3} set \cite{Ball:2012cx} while keeping the default values in the MC tools. We interfaced the truth level events to {\tt CheckMATE\,2.0.2} \cite{Dercks:2016npn,Kim:2015wza,Drees:2013wra,Tattersall:2016nnx} which is based on the fast detector simulation {\tt Delphes\,3.4.0} \cite{deFavereau:2013fsa} and the jet reconstruction tool {\tt Fastjet\,3.2.1} \cite{Cacciari:2011ma}. {\tt CheckMATE} tests a model point against current ATLAS and CMS searches at $95\%$ confidence level.\footnote{{\tt CheckMATE} contains a number of high luminosity searches which allows the user to estimate the potential discovery reach, e.g. \cite{Kim:2016rsd}. In addition, with the help of the AnalysisManager new search strategies can be implemented \cite{Rolbiecki:2015lsa,Kim:2014noa}.} We take the leading order cross section from {\tt Pythia\,8} and normalise it with a conservative $\mathcal{K}$ factor of 20\%. We do not generate MC events with at least one additional parton at matrix element level and matched with the $p_T$ ordered parton shower of {\tt Pythia\,8}. In principle, our approach might be problematic for compressed spectra \cite{Dreiner:2012gx,Dreiner:2012sh,Drees:2012dd}. However, in this letter the effects are negligible, since the decay of the NLSP into the gravitino will dominate the final state topology.

We have performed several grid scans in the softbreaking parameters of the electroweakino sector as well as $\mu$ with the SM-like Higgs mass fixed to $m_h=125$ GeV. In Table~\ref{tab:scan_parameters}, we list all scenarios and the scan parameters. They will be discussed in more detail further below. For each grid point, $5\cdot 10^4$ events have been generated.

We test each grid point against all relevant ATLAS and CMS searches implemented into {\tt CheckMATE}. % {as well as the two most relevant CMS Run-2 searches}. 
The most relevant searches of which are summarised in Table~\ref{tab:lhc_searches8tev}. 
%The implementation of Ref.~\cite{Aad:2015hea} has 
The implementations of Refs.~\cite{Aad:2015hea,CMS:2017vdc} into {\tt CheckMATE} have % CMS:2017mwo,
been done for this project and are now publicly available within the standard {\tt CheckMATE} repository.
All listed analyses are fully validated, and validation information is provided on the official webpage\footnote{\url{https://checkmate.hepforge.org/}} for the interested reader.

Every analysis covers a large number of signal regions which provides sensitivity to a vast range of mass hierarchies and final state multiplicities. For each search, the {\it best signal region} corresponds to the signal region with the best {\it expected} exclusion potential. This approach is again followed to choose the {\it best search}. As a result, the best observed limit is not always used to determine the limit and thus naively, the limits might be weaker. However, on the other hand, our limits are insensitive to downward fluctuations in the data which we would expect taking into account the huge number of signal regions. We select the {\it best search} and then compare our estimate of number of signal events with the observed limit at 95\% confidence level \cite{Read:2002hq},

\begin{equation}
r=\frac{S-1.96\cdot\Delta S}{S_{\rm exp}^{95}}\,,
\label{eq:r}
\end{equation}

where $S$, $\Delta S$, and $S_{\rm exp}^{95}$ denotes the number of signal events, the uncertainty due to MC errors and the 95$\%$ confidence level limit on the number of signal events, respectively. We only consider the statistical uncertainty due to the finite Monte Carlo sample with $\Delta S=\sqrt{S}$. The $r$ value is only calculated for the expected best signal region. 
Note that a combination of signal regions is currently not yet possible in {\tt CheckMATE}, therefore the limits we derive in the following are conservative.
A model point is excluded if $r$ is larger than one. However, here we define a point as allowed for $r<0.67$ and excluded for $r>1.5$. Benchmark points which fall in the region $0.67<r<1.5$ might be excluded or allowed but due to missing higher order corrections and other systematic errors, we do not classify them as excluded or allowed.

\begin{table}[tbh]
\begin{center}
\begin{tabular}{l|l|l|l}
$\sqrt{s}$ & Reference & Final State & $\mathcal{L}$\,[fb$^{-1}$]\\
\hline
\hline
8\,TeV & 1507.05493 (ATLAS) \cite{Aad:2015hea} & $1(2)\gamma+0(1)\ell$ & 20.3\\
& & +(b)-jets+$\met$ &\\
& 1411.1559 (ATLAS) \cite{Aad:2014tda} & $\gamma+\met$ & 20.3 \\
& 1501.07110 (ATLAS) \cite{Aad:2015jqa} & $h+\met$ & 20.3 \\
& 1403.4853 (ATLAS) \cite{Aad:2014qaa} & 2$\ell$+$\met$& 20.3 \\
& 1404.2500 (ATLAS) \cite{Aad:2014pda} & SS 2$\ell$ or 3$\ell$ & 20.3\\
& 1405.7875 (ATLAS) \cite{Aad:2014wea} & jets + $\met$ & 20.3\\
& 1407.0583 (ATLAS) \cite{Aad:2014kra} & 1$\ell$+($b$) jets+$\met$ & 20.0\\
& 1402.7029 (ATLAS)  \cite{Aad:2014nua} & 3$\ell$+$\met$& 20.3 \\
& 1501.03555 (ATLAS) \cite{Aad:2015mia} & 1$\ell$+jets+$\met$ & 20.3\\
& 1405.7570 (CMS) \cite{Khachatryan:2014qwa} & 1,\,SS-OS2,\,3,\,4$\ell$& 20.3 \\
&                                            & $+\met$ & \\
\hline
13\,TeV & %CMS-PAS-SUS-16-44 
1709.04896 (CMS) 
\cite{CMS:2017mwo,Sirunyan:2017obz} & $2\,h+\met$ & 35.9 \\
 & CMS-PAS-SUS-16-46 \cite{CMS:2017vdc} & $\gamma + \met$ &  35.9\\
 & 	ATLAS-CONF-2017-039 \cite{ATLAS:2017uun} & $2-3\, \ell$ & 36.1\\
 & 1704.03848 (ATLAS) \cite{Aaboud:2017dor} & high-en. $\gamma+\met$ & 36.1\\
 & ATLAS-CONF-2016-096 \cite{ATLAS:2016uwq} & $2-3\,\ell+\met$ & 13.3\\
 & 1604.01306 (ATLAS) \cite{Aaboud:2016uro} & $\gamma + \met$ & 3.2
\end{tabular}
\end{center}
\caption{Summary of the most relevant analyses. In addition, we have also taken into account all other searches implemented in {\tt CheckMATE} for our study.
If available, the analyses are referenced by their arXiv number, otherwise by their internal report number. The third column denotes the final
state topology, and the fourth column shows the total integrated luminosity.}
\label{tab:lhc_searches8tev}
\end{table}

Before presenting our numerical results, we first want to provide a rough overview of the relevant Run 1 and Run 2 searches and their current status. ATLAS and CMS have presented a vast number of studies covering a large selection of final state topologies. We want to investigate the impact on our GGM scenario by taking into account all relevant 8 TeV and 13 TeV searches since only a few dedicated GGM studies have been on the market so far. In Table \ref{tab:lhc_searches8tev} we list all relevant searches implemented in {\tt CheckMATE}, divided into analyses performed on 8\,TeV and 13\,TeV data. 

\subsection{8\,TeV analyses}
\label{subsec:8TeV}
The search \cite{Aad:2015hea} is a tailor made study targeting GGM inspired models and provides the backbone of our suit of LHC analyses. The authors search for events with high energetic isolated photons and large transverse momentum in the full data set of Run 1. The analysis contains ten signal regions which can roughly be divided into four classes. The first class demands diphoton final states with large transverse momentum targeting gluino and wino pair production channels with subsequent decays into binolike NLSPs. The remaining signal regions only require a single high energetic photon in the event topology which are further classified in ($b$)-jet multiplicity signal regions. This class of search channels probes scenarios with gluinos and higgsinos and subsequent cascade decays into bino NLSPs. Finally, signal regions requiring an isolated electron or muon in the final state are designed to focus on wino NLSP scenarios. Ref.~\cite{Aad:2014tda} is designed to look for events with a single large transverse momentum photon balancing large missing transverse energy. 
The search targets models of simplified dark matter where pair-produced dark matter recoils against a photon of SUSY-inspired scenarios with compressed spectra where pair produced squarks are produced in association with a photon. The SM backgrounds of the mono-photon signal can be suppressed by demanding a high momentum photon and large missing transverse momentum while requiring a strict lepton veto. As we will see, Ref.~\cite{Aad:2015jqa} performs pretty well in certain regions of parameter space. This analysis searches for chargino--neutralino pairs decaying into a $W$ boson and Higgs boson final state. Here, the authors exploit the emerging lepton of the $W$ boson decay as well as the diphoton, bottom pairs and the $WW^*$ final state of the Higgs decay to isolate the signal from the countless SM events. We also employed many other SUSY searches which might be sensitive to our GGM scenario such as multi--lepton searches \cite{Aad:2014qaa,Aad:2014nua} and studies covering multi jet plus lepton final states \cite{Aad:2014kra}. 

\subsection{13\,TeV analyses}
\label{subsec:13TeV}

Most 13 TeV analyses are designed to maximise the sensitivity in the high mass region, and  GGM-focused searches are quite rare. This is in particular the case for ATLAS for which {\tt CheckMATE} is optimized, as is also represented in the catalogue of implemented searches. Ref.~\cite{ATLASCollaboration:2016wlb} searches for two photons and large transverse momentum and they interpret their results in GGM models. Its signal regions resemble the diphoton signal regions of the 8 TeV study \cite{Aad:2015hea} aiming at very heavy gluinos cascade decaying into diphoton final states. Due to the heavy masses involved, they can demand a very tight cut on the effective mass of $m_{\rm eff}>1.5\,$TeV to separate the diphoton signal from the large SM backgrounds. This analysis is therefore not sensitive to the comparably small cross sections we are dealing with here. The other search is a conference note~\cite{ATLAS:2016fks} looking for photons, jets and large missing transverse momentum in the final state. Again, they demand a very large cut $m_{\rm eff}>2\,$TeV. \
%MKout{As a consequence only 8 TeV analyses are relevant for us and we will proceed with the 8 TeV searches and present our numerical results in the next section.}
The situation is different for CMS where two analyses look for electroweakino production in a GMSB context. 
Ref.~\cite{CMS:2017mwo,Sirunyan:2017obz} looks for higgsino pair production with the subsequent decay into $h\tilde G$ each. %, whereas only the $b\bar b$ channel of the Higgs decay is taken into account. 
%Signal events must feature $p_T^{\rm miss}>150\,$GeV, i.e. it is designed for 
%higgsinos with masses of several hundreds of GeV.
%In addition, 4-5 jets of which at least two pass tight $b$-tagging requirements are required. Leptons are vetoed against.
Signal events must feature $p_T^{\rm miss}>150\,$GeV, 4-5 jets out of which two pass tight $b$-tagging requirements, and no leptons.
The data is interpreted within a simplified model assuming that 100\%  of the $\tilde \chi^0_1$ decay into $h\tilde G$ as well as that 
%In addition it is assumed that 
both $\tilde h^0_2$ and $\tilde h^\pm$ decay into $\tilde h^0_1$, i.e. that each produced higgino results in a Higgs final state.
Neither the former nor the latter is usual in a realistic GMSB model,  see for instance Eqs.~(\ref{eq:partialwidths}) and (\ref{eq:kappas}). In fact, depending on $\tan\beta$ and sign($\mu$), typically the $Z\tilde G$ final state is much more abundant, and only for small $\tan\beta$ and negative $\mu$, a branching ratio of $\tilde h^0_1 \to h \tilde G$ of up to $\sim 80\%$ is possible.\footnote{This issue is considered in the CMS electroweakiono combination \cite{CMS:2017sqn} where the higgsino mass bound is shown as a function of the branching fraction into $h\tilde G$.}
The charged higgsino state, in turn, is only slightly heavier than $\tilde \chi^0_1$ so that the decays into the latter are phase-space suppressed. Depending on the gravitino couplings to matter (and therefore its mass), $\tilde h^\pm$ will almost exclusively decay into $W^\pm \tilde G$ if $m_{\tilde G}$ is small ($\mathcal O({\rm eV})$ or less) or dominantly into $\tilde \chi^0_1 q_i\bar q_j$ if $m_{\tilde G}$ is larger ($\mathcal O(100\,{\rm eV})$ or more). As the production cross section of higgsinos is dominated by the associated production of $\tilde h^\pm \tilde h^0_i $, this means that the actual rates of $\sigma(4 b)$ through higgsino production are reduced by many orders of magnitude if $m_{\tilde G}\lesssim 1\,$eV, making the search insensitive, or only a factor of $\mathcal O(2-10)$ if $m_{\tilde G}$ is larger.\footnote{Note that prompt NLSP decays are still possible for gravitino masses of $\mathcal O(100\,{\rm eV})$.} 
% Unfortunately, this makes the search \cite{CMS:2017mwo,Sirunyan:2017obz} insensitive to higgsino production in a realistic model with the currently-acquired data.\footnote{See e.g. Fig.~9 in Ref.~\cite{Sirunyan:2017obz}: only if the rescaling between the full and the simplified model amounts to a factor $\gtrsim 0.4$ would this search be sensitive. With the situation at hand, i.e. orders of magnitude difference, the theory cross-section curve is well below the current observed limit. }
This strong dependence on the model parameters is in contrast to the 8~TeV search \cite{Aad:2015jqa} which assumes a more realistic setting, only requiring one Higgs in the final state.

%\MK{\bf Maybe cut down this long description a bit?}

Although we have implemented \cite{CMS:2017mwo,Sirunyan:2017obz} into {\tt CheckMATE}, we will not make use of this implementation in what follows. The reason is that the $b$-tagging efficiency in CMS cannot be modelled accurately enough in {\tt CheckMATE} to date.\footnote{In {\tt CheckMATE}, it is currently only possible to define one $b$-tagging efficiency curve. The CMS analysis, however, uses several working points of very different shapes. An averaging over these curves did not satisfactorily reproduce the results of \cite{CMS:2017mwo,Sirunyan:2017obz}.}
Therefore, we estimate the bounds coming from this search by rescaling the production cross sections by the respective higgsino properties and compare them directly to the observed exclusion obtained in \cite{Sirunyan:2017obz}.

% Hence, a re-interpretation as done below is necessary for obtaining the bounds on a full model. {\bf MK: check if the following statement is true:} \emph{Consequently, both the jet requirement as well as the lepton veto hinder this search to be much more sensitive than the 8\,TeV analyses as we will see.}

Ref.~\cite{CMS:2017vdc} is designed for GMSB scenarios with at least one NLSP decay into $\gamma \tilde G$. Accordingly, it looks for at least one high-$p_T$ photon together with at least 300\,GeV of $\met$. The analysis is divided into four different signal regions depending on the value of $S_T^\gamma = p_T^{\rm miss} + \sum_{\gamma_i} p_T (\gamma_i)$, with a lower cut of 600\,GeV. Beside coloured production mechanisms, the results are also interpreted in terms of wino-like neutralino-chargino production, assuming $\tilde \chi_1^\pm \to W^\pm \tilde G$ and $\tilde \chi^0_1 \to \gamma \tilde G$ with 100\% branching ratio each. Also here, the results have to be confronted with a more realistic scenario in which a neutral wino is much more likely to decay into $Z\tilde G$ instead. Unlike the comparison to the 8\,TeV search of Ref.~\cite{Aad:2015hea} (which assumes a realistic scenario), we therefore expect a significant loss of experimental reach when comparing our scenarios with this analysis.

In addition to these searches, we further test against other, non-dedicated searches. Those include two dark matter-inspired searches from ATLAS looking for a photon and $\met$ \cite{Aaboud:2016uro,Aaboud:2017dor}, which are similar to the 8\,TeV analysis in Ref.~\cite{Aad:2014tda}. Note that  Ref.~\cite{Aaboud:2017dor} uses large thresholds for both the photon as well as the missing transverse momentum of 150\,GeV each. In addition, we include searches for two or three leptons in the context of `conventional' (i.e. without gauge mediation) electroweakino production and decay \cite{ATLAS:2016uwq,ATLAS:2017uun}.

\section{Numerical Results}
\label{sec:results}

In order to study the phenomenology of the GGM models we focus on a set of scenarios reflecting all possible mass hierarchies in the electroweakino sector while assuming that the gravitino is always the LSP. As discussed before we have two input parameters in the gaugino soft breaking sector namely $M_1$ and $M_2$ as well as $\mu$ and $\tan\beta$. We consider scenarios with bino, wino, higgsino NLSP scenarios as well as mixed eigenstates NLSP candidates. We explicitly checked that our limits on the model parameters are barely sensitive to the exact value of $\tan\beta$ and thus we fix $\tan\beta=10$ in the remainder of our letter. In the case where differences appear for other choices, we discuss them and show the respective results. In the following, we perform several two-dimensional grid scans while fixing the remaining input parameters. We have chosen four scenarios which are summarised in Table \ref{tab:scan_parameters}. These four scenarios are sufficient to consider all possible mass orderings in the electroweakino sector with a gravitino LSP. In our first scenario {\bf I} we concentrate on light binos and winos while decoupling the higgsino sector. Since the production cross section for purely bino-like eigenstates is negligible, the signal rate is governed by the production channels of electroweakinos with large wino composition. Analogously in scenario {\bf II}, we consider light binos and higgsinos while the winos are decoupled. Similar to scenario {\bf I} only higgsino-like final states have a sufficiently large cross section which can be probed at the LHC. In scenario {\bf III}, we take a closer look at scenarios with simultaneous presence of light higgsinos and winos. In addition, we assume a light bino. The presence of a light bino with a mass of 50 GeV in the mass spectrum warrants a bino NLSP. Finally, in scenario {\bf IV} we consider higgsinos and winos which can be kinematically accessed at the LHC. However, we factor out the impact of the bino on the collider phenomenology by setting its mass to 2 TeV.

\begin{table}
\centering
\begin{tabular}{c|c|c|c|c} %\hline \hline %\toprule
Scenario & $M_1$ [GeV] & $M_2$ [GeV] & $\mu$ [GeV] & Description \\ \hline\hline
{\bf I} & [100, 1000] & [100, 1000] & 2000 & $\mu$ decoupled \\
{\bf II} & [100, 1000]  & 2000 & [100, 1000] & $M_2$ decoupled\\
{\bf IIb} & [100, 1000]  & 2000 & [-1000, -100] & $M_2$ decoupled\\
{\bf III} & 50 & [100, 1000] & [100, 1000] & light bino\\
{\bf IV} & 2000 & [100, 700] & [100, 700] & heavy bino
%\\ \hline\hline %\bottomrule
\end{tabular}
\caption{For all scenarios, we effectively decouple the scalar sector, the gluinos as well as pseudoscalars from the spectrum while keeping the Higgs mass to 125 GeV. We use $\tan\beta=10$ throughout except for scenario {\bf IIb} where we set $\tan\beta=3$.
}\label{tab:scan_parameters}
\end{table}

\subsection*{Scenario {\bf I}}

\begin{figure}[ht]
	\begin{center}
        \includegraphics[width=.85\linewidth]{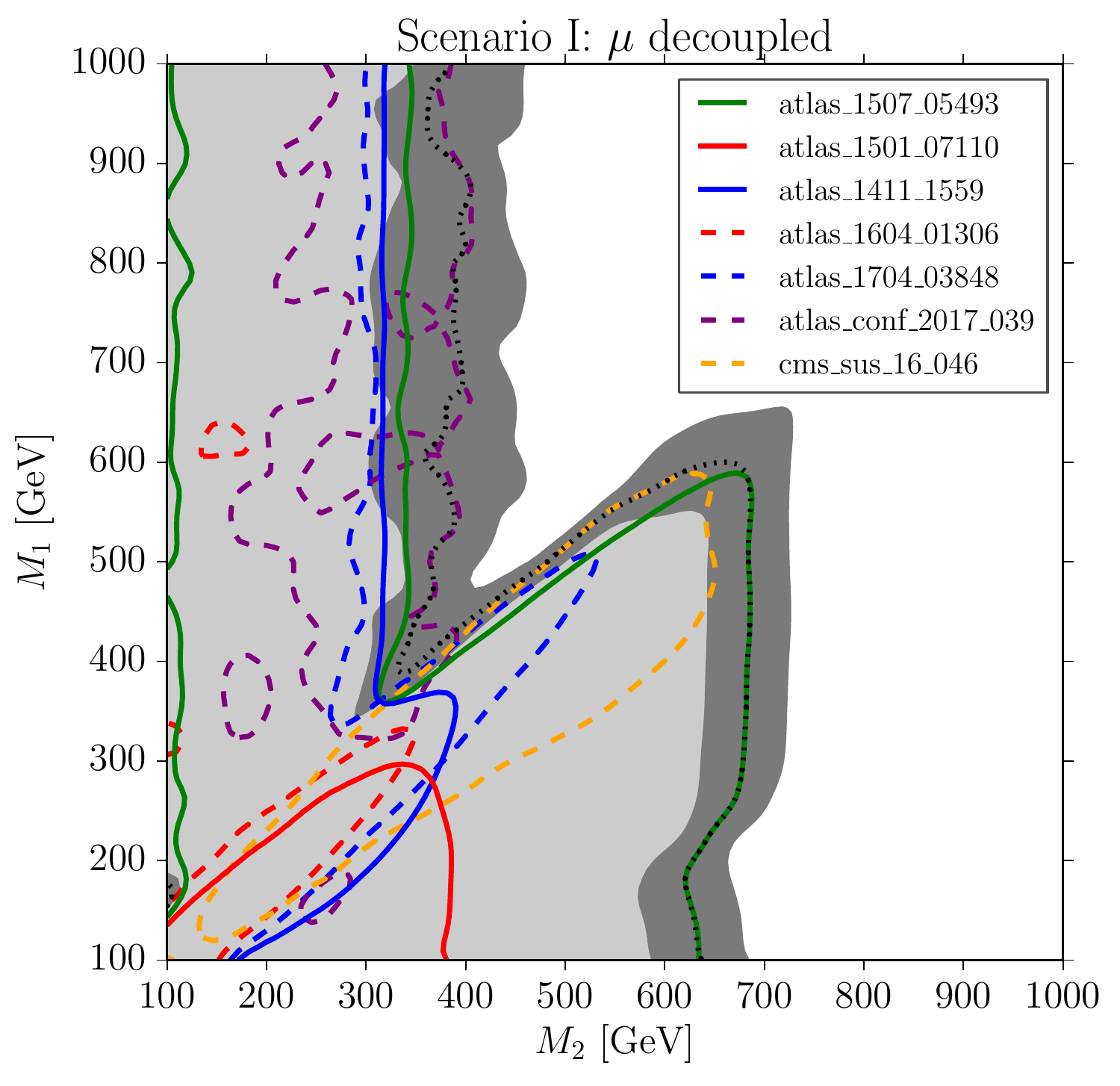}%{analysis_mudecoupled.pdf}
	\end{center}
\caption{95\% confidence level limit on the $M_1$ -- $M_2$ plane from ATLAS and CMS searches performed at the center of mass energy $\sqrt{s}=8$ TeV. The green, blue and red full lines correspond to 95\% C.L. limits from recasting the 8\,TeV analyses of Refs.~\cite{Aad:2015hea}, \cite{Aad:2014tda} and \cite{Aad:2015jqa}. %The black curve corresponds to the envelope around the best exclusion limit. The light gray area is excluded from these analyses while the white area is still allowed parameter space.
The constraints from the 13\,TeV searches are displayed in red \cite{Aaboud:2016uro}, blue \cite{Aaboud:2017dor}, purple \cite{ATLAS:2017uun} and orange \cite{CMS:2017vdc} dashed lines.
For the best exclusion line, we show in dark gray the estimated uncertainty bands. 
We also show in black %dashed 
dotted the best exclusion line as provided by {\tt CheckMATE} when considering all implemented searches. Due to the slightly different approach in the limit setting explained in \cref{sec:recasting}, there can appear small differences with respect to the envelope of the coloured exclusion lines.
Here we decoupled the higgsino sector by fixing $\mu=2\,$TeV. }
\label{fig:anal_mudecoupled}
\end{figure}

In Fig.~\ref{fig:anal_mudecoupled}, we show the best exclusion limit at 95\% confidence level denoted by the black dashed curve in the  $M_1$--$M_2$ plane while $\mu$ is fixed to 2 TeV. The dark gray shaded area corresponds to our theoretical uncertainty while the light gray one represents the excluded region. We also present the exclusion potential of the three  most important 8\,TeV searches in green \cite{Aad:2015hea}, blue \cite{Aad:2014tda} and red \cite{Aad:2015jqa} full lines and of the four most important 13\,TeV searches in red \cite{Aaboud:2016uro}, blue \cite{Aaboud:2017dor}, purple \cite{ATLAS:2017uun} and orange \cite{CMS:2017vdc} dashed lines. 
Of the 8\,TeV searches,  Ref.~\cite{Aad:2015hea} provides the best sensitivity which is hardly surprising since this search is targeting GGM scenarios with bino and wino NLSPs. In particular, the signal regions optimised to the diphoton states are powerful. Here, the kinematical selection cuts are optimized to isolate events with wino production channels with subsequent decays into binos. They demand at least two photons with $p_T\ge75$ GeV each and $\met\ge150\,(200)$ GeV for low (high) mass bino NLSPs. The latter cut allows to be sensitive to compressed mass spectra as well as to scenarios with large mass splitting between the wino and the bino NLSP. Our results nicely agree with the results in Ref.~\cite{Aad:2015hea}. The experimental analysis presents mass limits in the wino--bino mass plane with a bino NLSP and their limit is relatively insensitive to the mass difference between wino and bino with the upper mass limit of roughly 700 GeV for the wino eigenstates which we are able to reproduce within the systematic uncertainty. Moreover, they present limits for a wino NLSP scenario. The most sensitive signal region requires at least one photon with $p_T\ge125$ GeV, $\met\ge120$ GeV and the presence of an isolated electron or muon. Here, they derive a limit of roughly 350 GeV on the wino mass. Again, we agree with their results within the theoretical uncertainty. The main contribution to the relevant signal comes from a production of a wino-like neutralino together with a wino-like chargino. While the photon comes from the neutralino decay, the isolated lepton comes from the direct chargino decay $\tilde \chi^\pm_1 \to W^\pm \tilde G$ and the corresponding leptonic $W$ decay.
The 13\,TeV GMSB analysis \cite{CMS:2017vdc} is less sensitive than Ref.~\cite{Aad:2015hea} almost throughout the plane, except that it cuts a little deeper into the region with both $M_1$ and $\mu$ large. The reason for the worse performance in the remaining area can be found in the comparatively large cuts on the photon momentum and the $\met$.

Fig.~\ref{fig:anal_mudecoupled} demonstrates that also Ref.~\cite{Aad:2014tda} performs very well in the wino NLSP region. The kinematical cuts on the transverse momentum of the photon and the missing transverse momentum are similar. However, they demand a lepton veto and thus no further cuts on kinematic quantities requiring a lepton are demanded. Nevertheless, similar limits are obtained compared to the GGM search of Ref.~\cite{Aad:2015hea}. 
Moreover, while Ref.~\cite{Aad:2015hea} becomes insensitive to low $M_2$ values  (see the green vertical line at $M_2\simeq 120\,$GeV), the usage of Ref.~\cite{Aad:2014tda} enables us to exclude also this low-mass region with LHC data.

Finally, in the region around $M_2 \sim 300\,$GeV and $M_1 > M_2$ (corresponding to a wino NLSP), the multilepton analysis \cite{ATLAS:2017uun} becomes sensitive. This is the case since the dominant decays of the neutral and charged winos are into $Z\gamma$ and $W\gamma$, respectively, leading to signatures of $3 \ell +\met$ as well as, more importantly, $2\ell + 2 j + \met$. Both types of signatures are probed by Ref.~\cite{ATLAS:2017uun}. Due to the large $\met$ cuts of at least 100\,GeV in the $2\ell +$jets signal regions, the search loses sensitivity below $\mu \sim 200\,$GeV.

\subsection*{Scenario {\bf II}}
\begin{figure}[ht]
	\begin{center}
        \includegraphics[width=.85\linewidth]{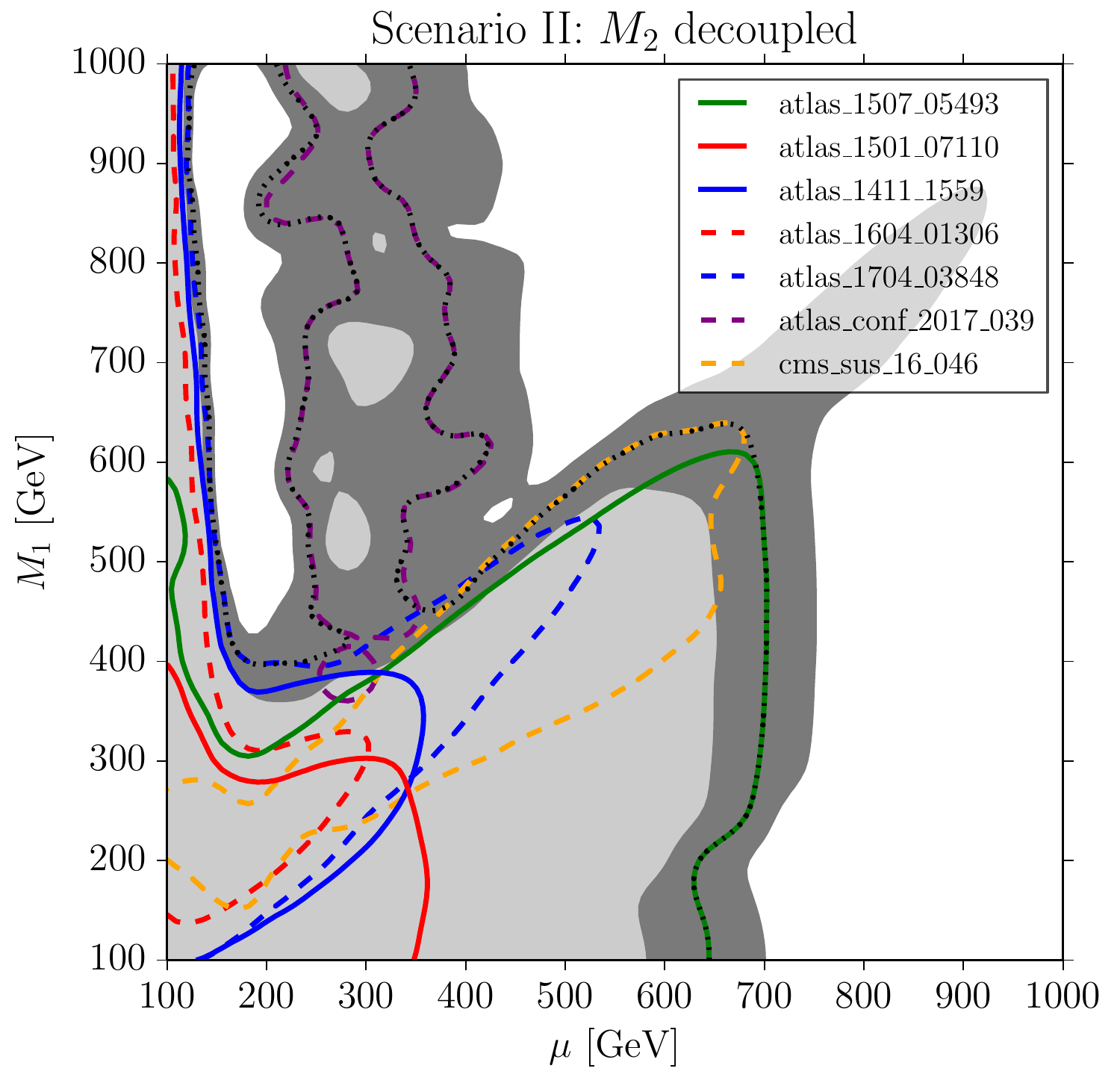}%{analysis_m2decoupled.pdf}
	\end{center}
\caption{Same as in \cref{fig:anal_mudecoupled} but in the $M_1$ -- $\mu$ plane using $M_2=2\,$TeV.}
\label{fig:anal_m2decoupled}
\end{figure}

We present our limits of scenario {\bf II} in Fig.~\ref{fig:anal_m2decoupled}. The shape of the excluded region is very similar to scenario {\bf I} for $\mu\ge300$ GeV. The diphoton signal regions in Ref.~\cite{Aad:2015hea} are relatively insensitive to the details of the decay products of the heavy neutralino states into the lightest neutralino and thus similar efficiencies in the signal regions are expected for wino and higgsino induced diphoton final states. The upper limit on the higgsino mass might be somewhat surprising since the production cross section for higgsino mass eigenstates are considerably smaller than for wino eigenstates. On the other hand, we have two higgsino eigenstates which are close in mass and therefore compensate for the smaller production cross section. 
As in Scenario~1, the GMSB-specific 13\,TeV analysis of Ref.~\cite{CMS:2017vdc} only adds mildly to the excluded region on the diagonal of the plot and becomes insensitive towards lower bino masses.

Benchmark points with higgsino NLSPs in the region $\mu\le300\,$GeV and $M_1\ge300\,$GeV are very weakly constrained by both Run 1 and 2 data. We can clearly see that the limits on the higgsino mass parameter are substantially smaller than in the wino NLSP case with a limit very close to the LEP II bound \cite{Agashe:2014kda}.  In the higgsino NLSP region,  the decay $\tilde \chi^0_1 \to \tilde G \gamma$ is only possible through the small bino admixture present in the NLSP state. Consequently, as soon as the bino mass is considerably larger than $\mu$, the photonic decay
is suppressed by the much larger branching ratio into $\tilde G Z$ and $\tilde G h$. Therefore,whenever the decay into $\tilde G h$ is kinematically accessible, BR($\tilde \chi^0_1 \to \tilde G \gamma$) ranges below a per-cent, making the signal efficiency in the diphoton signal regions negligible. 
The enhancement of the $Z$ final states leads to possible multi-lepton signatures which are captured by searches like Ref.~\cite{ATLAS:2017uun} as seen in the purple dashed lines. 
The corresponding enclosed area is analogous to the wino NLSP case of Fig.~\ref{fig:anal_mudecoupled}
%Ref.~\cite{ATLAS:2017uun} also considers final states with two leptons, jets and $\met$, which are particularly sensitive to the higgsino LSP case -- however, due to the large $\met$ cuts of at least 100\,GeV, the search is not sensitive below $\mu \sim 200\,$GeV. 
There are, however, two effects through which the single-photon searches of Refs.~\cite{Aad:2015hea,Aad:2014tda} become sensitive at least in the small-$\mu$ region. Firstly, 
%\MKout{However,} 
due to the small mass splitting between the lighter and the next heavier higgsino state, the %\MKout{two-body} 
decay of $\tilde \chi^0_2$ down to the NLSP 
%\MKout{into $\tilde \chi^0_1 \gamma$ is typically at the level of a few per-cent and} 
is mainly into a three-body final state %\MKout{otherwise} 
but also a significant fraction of typically a few per-cent decays into $\tilde \chi^0_1 \gamma$. 
Secondly, for small enough $\mu$, the decay mode $\tilde G h$ becomes kinematically suppressed, leading to a relative enhancement of the photonic mode $\tilde G \gamma$.
%\MKout{Through this extra photon from the NNLSP decay, the single photon signal regions of Refs.~\cite{Aad:2015hea,Aad:2014tda} are sensitive in a small window with low higgsino masses $\mu \lesssim 150~$GeV.}
We further see that, with increasing $M_1$, corresponding to a decreasing bino admixture within $\tilde \chi^0_1$, the experimental searches become insensitive to the higgsino NLSP scenario accordingly. 
In the region with both low $\mu$ and low $M_1$ where the NLSP is made of a significant bino-higgsino mixture, the mono-photon search of Ref.~\cite{Aad:2014tda} is actually providing the best 8\,TeV limits, and similarly Ref.~\cite{Aaboud:2017dor} with the best 13\,TeV limits. In this region, the NLSP decay into $\tilde G \gamma$ ranges around 10\,\% so that the analysis is quite sensitive to a pair of neutralinos where one decays into $\tilde G \gamma$ and the other into $\tilde G$ and two jets or two neutrinos from the $Z/h$ decay.

\begin{figure}[ht]
	\begin{center}
        \includegraphics[width=.85\linewidth]{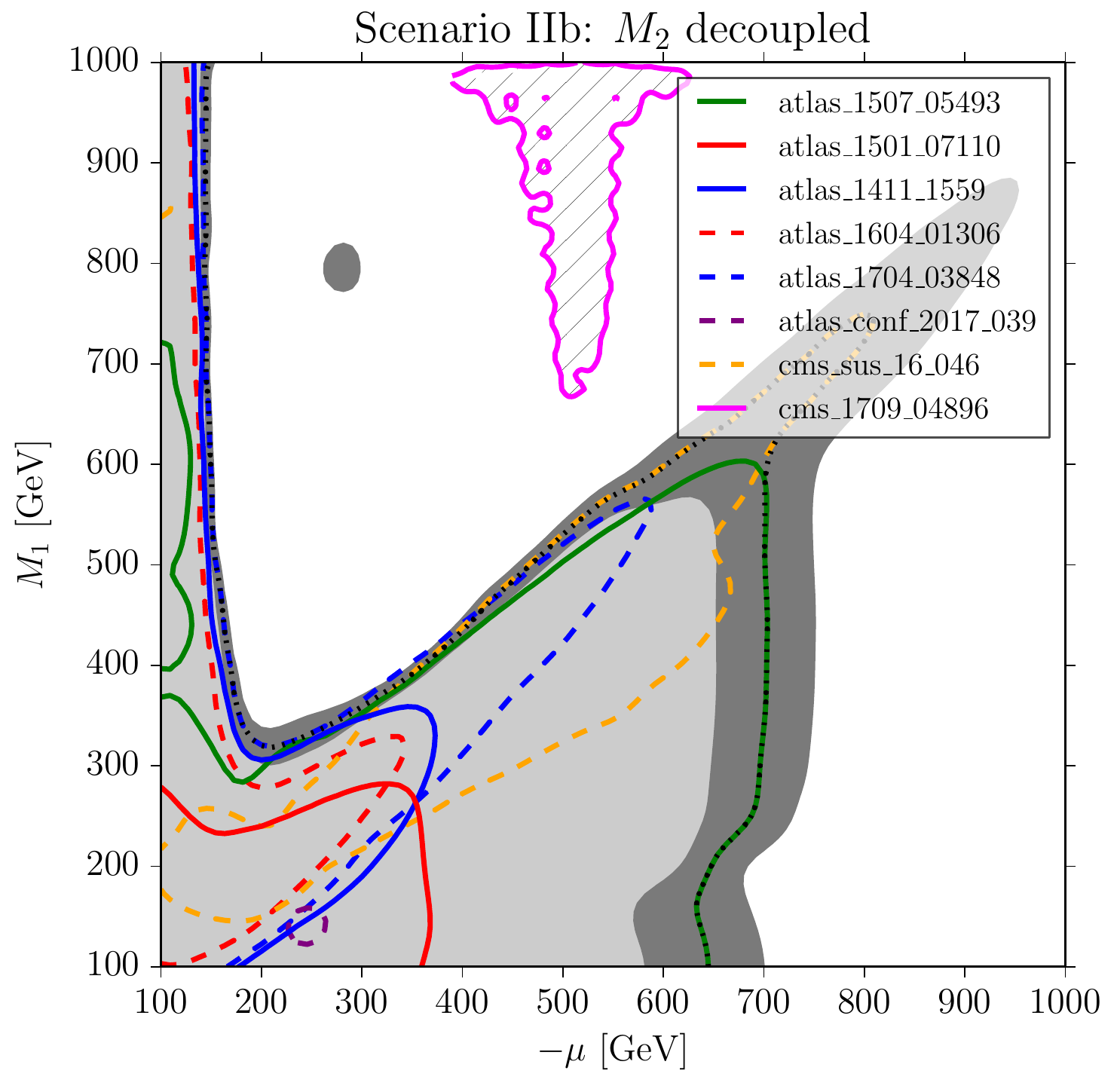}%{analysis_m2decoupledtb3neg.pdf}
	\end{center}
\caption{Analogue of \cref{fig:anal_m2decoupled} but using $\tan\beta =3$ and $\mu<0$. In addition we show in the magenta contour and the hatched region the projected limits from Ref.~\cite{Sirunyan:2017obz}, assuming a higher gravitino mass of $m_{\tilde G}=100\,$eV.}
\label{fig:anal_m2decoupledtb3neg}
\end{figure}

Finally note also that the decay properties of the higgsino are quite sensitive to the exact value of $\tan \beta$ and also the sign of $\mu$: if $\tan\beta$ is small, then using \cref{eq:kappas}, the partial decay width of the higgsino into $h\tilde G$ is enhanced if $\mu<0$. Correspondingly, both the branching ratios into $\tilde G Z$ and $\tilde G \gamma$ are suppressed. 
This scenario is depicted in \cref{fig:anal_m2decoupledtb3neg} where we repeat the scan of \cref{fig:anal_m2decoupled} with $\tan\beta=3$ and $\mu<0$ while keeping the Higgs mass at the observed value. One can clearly see that the above-mentioned region with significant higgsino-bino mixing is less constrained by existing searches, in particular Ref.~\cite{Aad:2014tda} looses sensitivity. In addition, the multilepton search \cite{ATLAS:2017uun} becomes inefficient because of the decreased branching ratio into $Z$ final states.
However, in the narrow band of small $\mu$ but large $M_1$, there are only very small differences in the exclusion bounds with respect to \cref{fig:anal_m2decoupled}.
%, which is due to the kinematical suppression of the $\tilde G h$ decay mode. 
The reason here is simply that the lightest higgsino is slightly lighter for $\mu<0$ than for $\mu>0$, so that the region of kinematical suppression of the $\tilde \chi^0_1 \to \tilde G h$ decay is already reached for slightly higher values of $\mu$.

Finally, we present in \cref{fig:anal_m2decoupledtb3neg} the projected limits from the di-Higgs search Ref.~\cite{Sirunyan:2017obz} in magenta, the hatched region being excluded by this analysis. Note that for this line we have used a different gravitino mass of 100\,eV which ensures that the decay $\tilde h^\pm \to \tilde h^0_1 q_i \bar q_j$ dominates over $\tilde h^\pm \to \tilde W^\pm \tilde G$, which is the decay which occurs in the so-far studied case of $m_{\tilde G}=1\,$GeV. Limits on the latter scenario do not exist because of the reduced Higgs abundance in the final state. As discussed in section~\ref{subsec:13TeV}, a proper recast with {\tt CheckMATE} is not possible, so that we projected the unfolded limits on cross section times branching fraction onto the scenario at hand. As expected, this analysis can only exclude rather pure higgsino states (i.e. the excluded region grows with decreasing bino fraction) in the regions of its best sensitivity, i.e. between roughly 400 and 600\,GeV. This is in contrast to the much stronger limits on the less realistic simplified model considered in Ref.~\cite{Sirunyan:2017obz} where higgsino masses between 230 and 770\,GeV have been excluded.

\subsection*{Scenario III}

%\begin{figure*}[htbp]
\begin{figure}[t]
	\begin{center}
           \includegraphics[width=.85\linewidth]{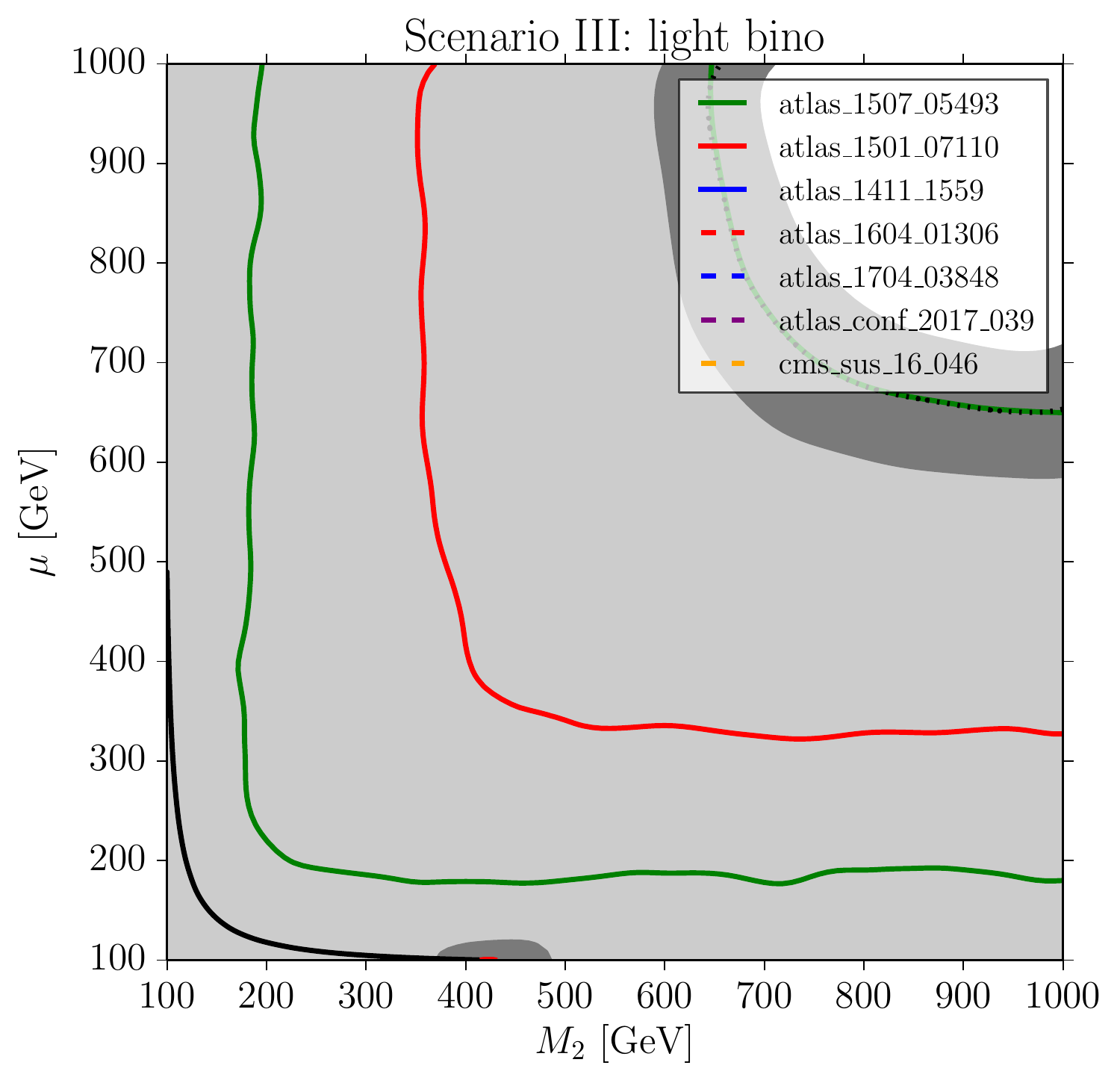}%{analysis_lightbinofixed.pdf}
	\end{center}
\caption{Same as in \cref{fig:anal_mudecoupled} but in the $M_2$ -- $\mu$ plane and $M_1=50\,$GeV.
Here we show in black also the LEP~II limit which excludes chargino masses below 104\,GeV \cite{Agashe:2014kda}.}
\label{fig:anal_lightbino}
\end{figure}

Now, we investigate how mixed wino--higgsino states can be constrained by Run-1 data. Fig.~\ref{fig:anal_lightbino} presents the excluded region in the $M_2$--$\mu$ plane. Here, we fix the bino mass to 50 GeV. The excluded region extends to roughly 600\,GeV for pure higgsinos, winos as well as for mixed states. Our result is consistent with Fig.~\ref{fig:anal_mudecoupled} and \ref{fig:anal_m2decoupled} of scenarios {\bf I} and {\bf II}. In the previous two subsections, we showed that the upper mass limit for higgsinos and winos with a bino NLSP is almost independent of the mass splitting since the search~\cite{Aad:2015hea} has two signal regions covering the low and high mass neutralino LSP range. We can observe that, as also seen in \cref{fig:anal_mudecoupled}, the search~\cite{Aad:2015hea} looses sensitivity in the very low mass region $M_2\approx 200\,$GeV or $\mu \approx 200\,$GeV since the signal regions are optimized for heavier masses. However, Ref.~\cite{Aad:2015jqa} shows sensitivity in the respective regions and is able to exclude the remaining parameter space for low $M_2$ and $\mu$. This search is optimized for neutralino--chargino pair production with the heavy neutralino decaying into a Higgs boson. In particular, the diphoton signal regions are very sensitive in this region of parameter space. The selections cuts require an isolated lepton and two photons with $p_T\ge25,35$ GeV. In addition, they demand a moderate cut on missing transverse momentum of 40 GeV. Finally, they apply cuts to enhance sensitivity to leptonically decaying $W$ bosons.

As is seen in Fig.~\ref{fig:anal_mudecoupled}, none of the 13\,TeV analyses constrains the light-bino scenario. This is due to the fact that the $\met$ signal in this case is heavily reduced compared to the previous scenarios whereas all relevant 13\,TeV analyses use strong  $\met$ cuts of $\mathcal O({\rm 100\,GeV})$ or higher.

\subsection*{Scenario IV}

\begin{figure}[t]
	\begin{center}
        \includegraphics[width=.85\linewidth]{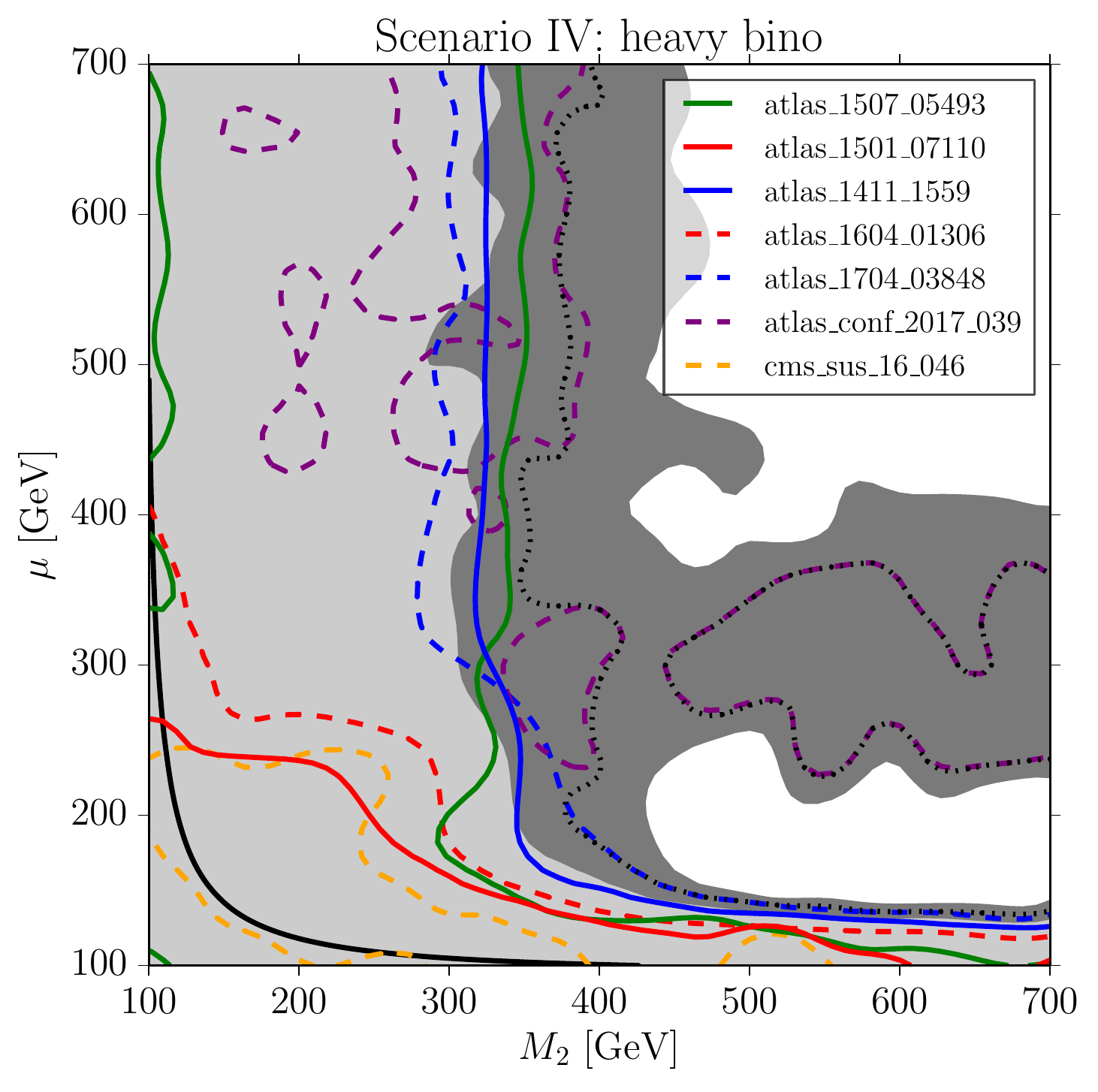}%{analysis_heavybino.pdf}
	\end{center}
\caption{Same as in \cref{fig:anal_lightbino} but using $M_1=2\,$TeV.}
\label{fig:anal_heavybino}
\end{figure}

In Fig.~\ref{fig:anal_heavybino}, we present the limits in the $M_2$ -- $\mu$ plane while decoupling the bino from the electroweakino mass spectrum. As expected, the limits are substantially weaker with the absence of the bino NLSP since the main  source of generating isolated photons is lost. Instead, while the wino still decays into $\tilde G\gamma$, cf. \cref{eq:partialwidths,eq:kappas}, the higgsinos decay into $\tilde G Z/h$ as also in the upper left corner of \cref{fig:anal_m2decoupled}.
The shape of the excluded region can easily be understood from our discussion of scenarios {\bf I} and {\bf II} and the results summarised in Figs.~\ref{fig:anal_mudecoupled} and \ref{fig:anal_m2decoupled}. We can read off the limits for decoupled higgsinos or winos in the region where the bino is not the NLSP candidate. Analogously to \cref{fig:anal_mudecoupled}, a wino NLSP is excluded up to $\sim 300\,$GeV. In the wino decoupled scenario, we observe that $\mu\lesssim 130\,$GeV is excluded, similar to \cref{fig:anal_m2decoupled}. 
Moreover, we can see that in the bino decoupled scenario, the GGM search \cite{Aad:2015hea} as well as the DM/compressed SUSY searches \cite{Aad:2014tda,Aaboud:2017dor} have comparable sensitivity while the electroweakino search \cite{Aad:2015jqa}, the GMSB analysis \cite{CMS:2017vdc} and the generic $\gamma+\met$ search \cite{Aaboud:2016uro} cover the low-mass region in the lower left corner of Fig.~\ref{fig:anal_heavybino}. 
As is also seen in scenario~{\bf{II}} in the mixed higgsino-bino region, the mixed higgsino-wino region in \cref{fig:anal_heavybino} is also best covered by the non-dedicated mono-photon searches \cite{Aad:2014tda,Aaboud:2017dor}.
%To summarise, SUSY searches not optimized for GGM scenarios can have similar sensitivity and even cover regions of parameter region missed by GGM searches. Our results clearly show the complementarity of searches.
Finally, as already visible in Figs.~\ref{fig:anal_mudecoupled}-\ref{fig:anal_m2decoupled}, the (also non-dedicated) multilepton analysis of Ref.~\cite{ATLAS:2017uun} is sensitive to the regions with wino as well as higgsino NLSPs with masses around 300\,GeV. As is seen in the dark grey shading, the results are not very conclusive yet for this analysis, however searches of this type will become important probes of these parameter regions once more data is accumulated.

%\begin{figure}[htbp]

\section{Conclusions}
\label{sec:conclusions}

In this letter we have presented limits on the electroweakino sector of different scenarios of general gauge mediation models. 
We have also demonstrated that although dedicated analyses which have been optimised for GGM models are very effective in constraining the parameter space, other searches which have not specifically been designed for this kind of scenarios are important as well and provide complementary constraints. In the case where $\mu$ is large and $M_1$ small we observe that $M_2$ is excluded for $M_2\lesssim 600$\,GeV, while this limit weakens to $M_2\lesssim 300$\,GeV when $M_1$ is large. We further observe  that in the case of light $M_1$ we can exclude $\mu \lesssim 600$\,GeV. If, however, the higgsino is lighter than both wino and bino, 
we find that dedicated GGM searches cannot exclude more parameter space than what is already ruled out by LEP~II. Including other experimental analyses as well, we can at least exclude higgsino masses smaller than the Higgs mass as long as the bino is not fully decoupled from the spectrum. Furthermore, multilepton analyses are now beginning to probe both bino and higgsino NLSPs with masses larger than about 200\,GeV.
Concluding, we have presented the current LHC bounds on pair-produced electroweakinos while highlighting and making use of the 
complementarity between both dedicated and non-dedicated analyses.

\vspace*{1cm}
\noindent{\bf \large Acknowledgments}

J.S.K would like to thank M. Drees and H. Dreiner and the University of Bonn for support and hospitality while this manuscript was prepared. M.E.K is supported by the DFG Research Unit 2239 ``New Physics at the LHC'' and thanks Giovanni Zevi Della Porta for interesting discussions. V.M.L acknowledges the support of
the Consolider-Ingenio 2010 programme under grant MULTIDARK CSD2009-00064, the European
Union under the ERC Advanced Grant SPLE under contract ERC-2012-ADG-20120216-
320421 and the BMBF under project 05H15PDCAA. V.M.L would like to thank Toby Opferkuch for valuable help and comments.
%----------------------
%BIBLIOGRAPHY
%----------------------

\small

\bibliography{lib}

\end{document}